\documentclass[10pt]{article}

\usepackage[letterpaper,margin=1in]{geometry}

\usepackage[sort]{natbib}
\usepackage{amsmath}
\usepackage{amsthm}
\usepackage{shortcutsg}
\usepackage{bm}
\usepackage{graphicx}
\usepackage{subcaption}
\usepackage[capitalise]{cleveref}
\Crefname{assumption}{Assumption}{Assumptions}

\theoremstyle{plain}

\theoremstyle{definition}

\usepackage{xcolor}

\title{Evaluating the Surrogate Index as a Decision-Making Tool Using 200 A/B Tests at Netflix}
\author{Vickie Zhang \and Michael Zhao \and Anh Le \and Maria Dimakopoulou \and Nathan Kallus}
\date{}

\newcommand{\textapprox}{\raisebox{0.5ex}{\texttildelow}}

\begin{document}
\maketitle

\begin{abstract}
Surrogate index approaches have recently become a popular method of estimating longer-term impact from shorter-term outcomes. In this paper, we leverage 1098 test arms from 200 A/B tests at Netflix to empirically investigate to what degree would decisions made using a surrogate index utilizing 14 days of data would align with those made using direct measurement of day 63 treatment effects. Focusing specifically on linear ``auto-surrogate" models that utilize the shorter-term observations of the long-term outcome of interest, we find that the statistical inferences that we would draw from using the surrogate index are \textapprox95\% consistent with those from directly measuring the long-term treatment effect. Moreover, when we restrict ourselves to the set of tests that would be ``launched" (i.e. positive and statistically significant) based on the 63-day directly measured treatment effects, we find that relying instead on the surrogate index achieves 79\% and 65\% recall. 
\end{abstract}

\section{Introduction}

For many firms operating digital platforms, A/B experimentation is a critical tool to guide product development \citep{gupta2019top} . When developing a new product, feature, or intervention, it’s common practice to run a short-term A/B test to determine whether it should be shipped. Although this short testing cycle enables faster iteration and expands the overall testing capacity, one major drawback is that long-term treatment effects cannot be directly measured. This can be especially problematic as it’s not uncommon for short- and long-term effects to deviate substantially due to phenomena such as novelty effects, user fatigue, and ``clickbaitness" among others \citep{kohavi2012trustworthy}.

Rather than rely on single short-term outcomes to proxy for long-term ones, surrogate index approaches \citep{athey2019surrogate} have become a popular way to harness rich observations to address this issue. This type of approach first builds a model to predict the long-term outcome of interest from a set of short-term outcomes. If this set of short-term outcomes taken together satisfies the ``surrogacy assumption" \citep{prentice1989surrogate}, namely that they fully mediate all of the treatment’s long-term effects and that their effect on the long term can be identified, then long-term treatment effects can be properly estimated using the predicted long-term outcome instead.

In this paper, we explore the efficacy of applying surrogate index methods at scale. We focus on what we term ``auto-surrogate" models, which predict outcomes from lagged versions of the outcome variable. Using data from 200 personalization algorithm A/B tests, we find that surrogate index models constructed using only 2 weeks of data would've led to the same product ship decisions \textapprox95\% of the time when compared to making a call based on 2 months of data. Comparing to what actually shipped if making the call at 2 months, we find the surrogate index approach achieves 79\% precision and 65\% recall. Under relatively reasonable assumptions, this level of recall suggests that the increased throughput of a shorter, faster testing cycle outweighs the increased reliability of directly measuring long-term treatment effects.  

\section{Setting}
Our analysis here leverages data from 1098 test arms from 200 A/B tests sampled from the personalization algorithms space. When we test personalization algorithms at Netflix, we typically employ single shot allocations, where we randomly sample and assign some fraction of our members into different treatment arms at approximately the same time. For our purposes, we’ll consider ``long-term" as 63 days (2 months) after allocation and focus on outcome metrics that average a per-day observation. 

For each A/B test that is run for at least 63 days post allocation, a simple difference-in-means produces direct and unbiased estimates of the long-term treatment effect of any test arm:

\begin{equation}
    \hat{\tau}_{a,63} = \frac{1}{63 |N_a|} \sum_{i \in N_a} \sum_{t=1}^{63} Y_{it} - \frac{1}{63 |N_0|} \sum_{i \in N_0} \sum_{t=1}^{63} Y_{it}
\end{equation}

\noindent Where $N_0$ denotes the set of members assigned to the control arm, $N_a$ denotes the set of members assigned to test arm $a$, and $Y_{it}$ denotes the observed value of the outcome of interest of member i on day t after allocation.

We compare these against estimates produced via a surrogate index. Specifically, we follow a similar setup to the empirical study of \citet{athey2019surrogate}, where we employ a linear auto-surrogate model specification where only the shorter-term observations of Y are used to construct the surrogate index. More formally:

\begin{equation}
    \mu_i = \beta_0 + \sum_{t=1}^{T} \beta_t Y_{it} + \epsilon_i
\end{equation}

\noindent Where $\mu_i = \frac{1}{63} \sum_{t=1}^{63} Y_{it}$ is the observed average daily outcome for user $i$ over the first 2 months and $T \leq 63$ denotes the ``order" of the auto-surrogate model, in this case, the number of days after initial allocation used to form the surrogate index (note that $T=63$  will produce identical estimates to the direct difference-in-means estimator above). 

After training this model, the long-term treatment effect can be estimated by computing the difference-in-means on the predicted values produced using the surrogate index: 

\begin{equation}
    \hat{\tau}_{a,T} = \frac{1}{|N_a|} \sum_{i \in N_a} \hat{\mu}_{i,T} - \frac{1}{|N_0|} \sum_{i \in N_0} \hat{\mu}_{i,T}
\end{equation}

\noindent Where $\hat{\mu}_{i,T} = \hat{\beta}_0 + \sum_{t=1}^{T} \hat{\beta}_t Y_{it}$, is the predicted average daily outcome for user $i$ produced using an auto-surrogate model of order $T$.

In addition to different choices of $T$, we also explore two different approaches to train the auto-surrogate model:
\begin{enumerate}
    \item Pre-Test: For a particular test, we train the model using the 63 days of data from the same set of users prior to their allocation:
    \item Similar Test: Train the model using data from another A/B test in the same product space 
\end{enumerate}

\begin{figure}
\centering
\begin{subfigure}{.6\textwidth}
  \centering
  \includegraphics[width=\linewidth]{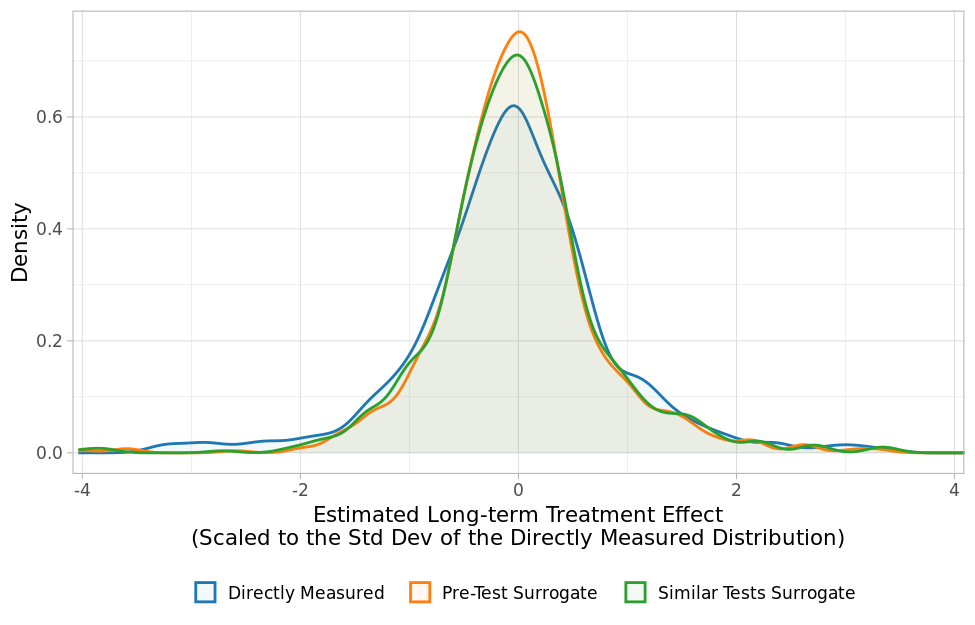}
  \label{fig:sub1}
\end{subfigure}%
\begin{subfigure}{.4\textwidth}
  \centering
  \includegraphics[width=\linewidth]{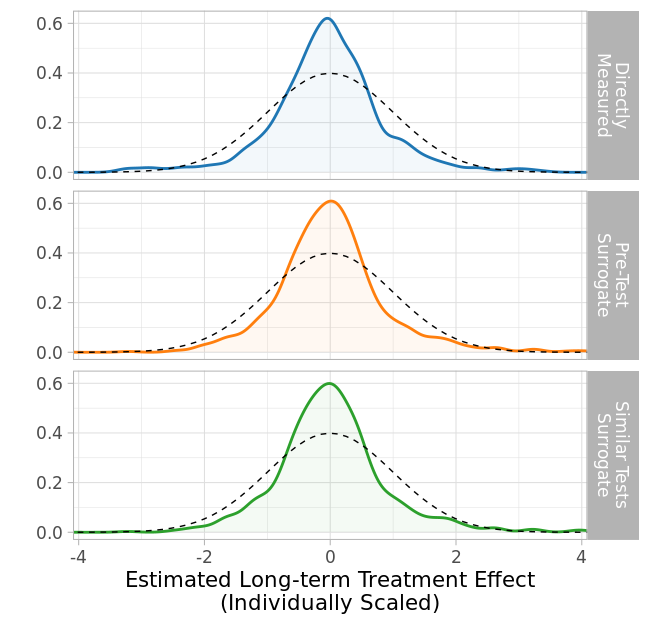}
  \label{fig:sub2}
\end{subfigure}
\caption{Density plots of the estimated long-term treatment effect produced by “Direct Measurement” (blue), “Pre-Test Surrogate Index” (orange), and “Similar Tests Surrogate Index” (green). The large figure on the left depicts the 3 distributions all scaled by the standard deviation of the directly measured distribution. The 3 vertically stacked figures on the right depict the densities individually scaled to their respective standard deviations, with the black dotted bell curve depicting the standard normal density for reference.}
\end{figure}

\noindent Figure 1 below shows scaled density plots of the long-term treatment effect estimates produced via direct measurement (blue) and the 2 approaches to training the auto-surrogate index. Consistent with the theory, we can see that both surrogate index estimators produce a distribution of estimated treatment effects that is lower in variance compared to direct measurement. Similar to what has been reported elsewhere \citep{peysakhovich2016combining, peysakhovich2018learning, coey2019improving, azevedo2020b}, we also find that our estimated treatment effects exhibit some degree of “fat-tailedness”.

\section{Results}

Figure 2 depicts the results from a single illustrative test, comparing pre-test and similar-test surrogate models with varying length of observation $T$ along with the running mean of observations (corresponding to a linear model where $\beta_0 = 0$ and $\beta_1 = \beta_2 = ... = \beta_T = \frac{1}{T}$). While representing only one test, Figure 2 suggests that a surrogate index can reliably recover the long-term treatment effect at much shorter timescale and with lower variance, consistent with what Athey et al. 2019 found. Moreover, we see that both the approaches to training are highly aligned with each other.

\begin{figure}
\centering
  \includegraphics[width=\linewidth]{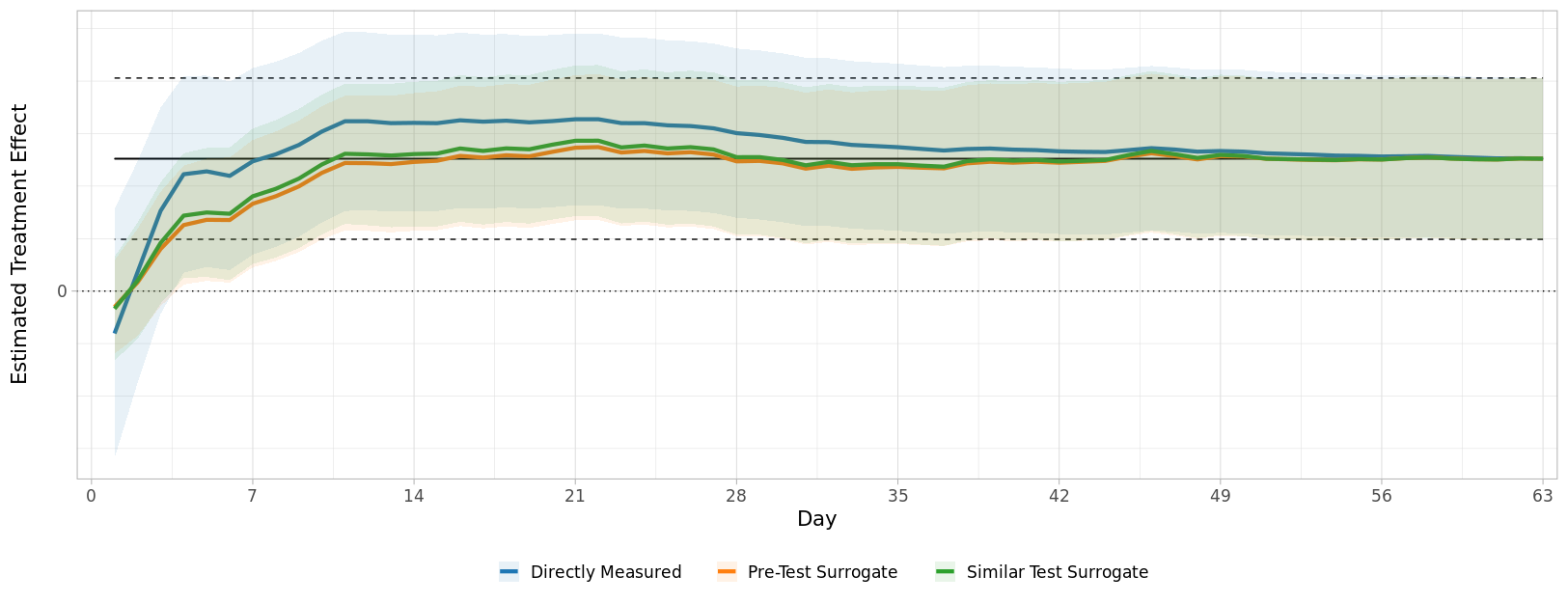}
\caption{Illustrating the estimated treatment effect for an example test conducted over a long period of time. Surrogate index trained on previous similar pre-test data (orange) and similar test data (green) utilizing $t$ days of data are compared with the directly measured treatment effect at day $t$ (blue). The solid and dashed black lines correspond to the directly measured 63-day estimated treatment effect and its 95\% confidence interval. The dotted black line represents 0 for reference.}
\end{figure}

Figure 3 compares the difference between the directly measured test read and those produced by the 2 surrogate indexes. In both cases, the expected difference is basically 0, suggesting that the surrogate index estimators unbiasedly recover the directly measured treatment effect (which is an unbiased estimator for the true treatment effect assuming SUTVA). Curiously however, it’s also worth noting that the distribution of these differences are also fatter-tailed compared to a normal distribution. 

\begin{figure}
\centering
  \includegraphics[width=\linewidth]{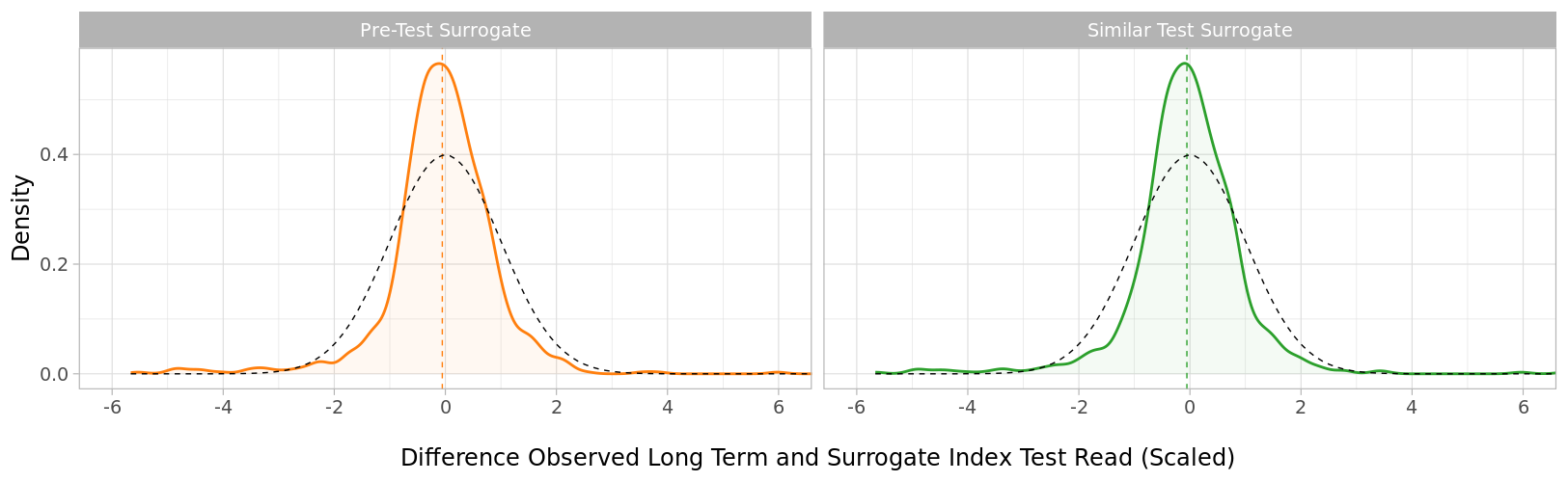}
\caption{Density plots of the differences between the Surrogate Index test read and the observed 2-month test read. The colored vertical lines denote the empirical means of these distributions. The black dotted bell shaped curve in both plots depicts a standard normal  for reference. Note that these are scaled by the standard deviation of the differences, so the scaling here is different from what is seen in figure 1 above.
}
\end{figure} 

Looking to Figure 4 shows that the surrogate index is highly aligned with the observed long-term test read in terms of statistical hypothesis testing agreeing \textapprox95\% of the time. Surprisingly, even though the surrogate index tends to have lower variance compared to the actual test read, it was more common for treatment estimates to be not statistically significant (\textapprox86.5\% for the surrogate index reads, and \textapprox79\% for the 2-month reads). Most crucially, supposing one would actually ship an intervention on a statistically significant positive read of long-term treatment effect, using the surrogate index achieves 79\% precision and 65\% recall with respect to the decision based on the 2-month read. Moreover, across our data, there was not a single case where surrogate index would've mistakenly concluded to launch an experience that was statistically negative when directly measured.

\begin{figure}
\centering
  \includegraphics[width=\linewidth]{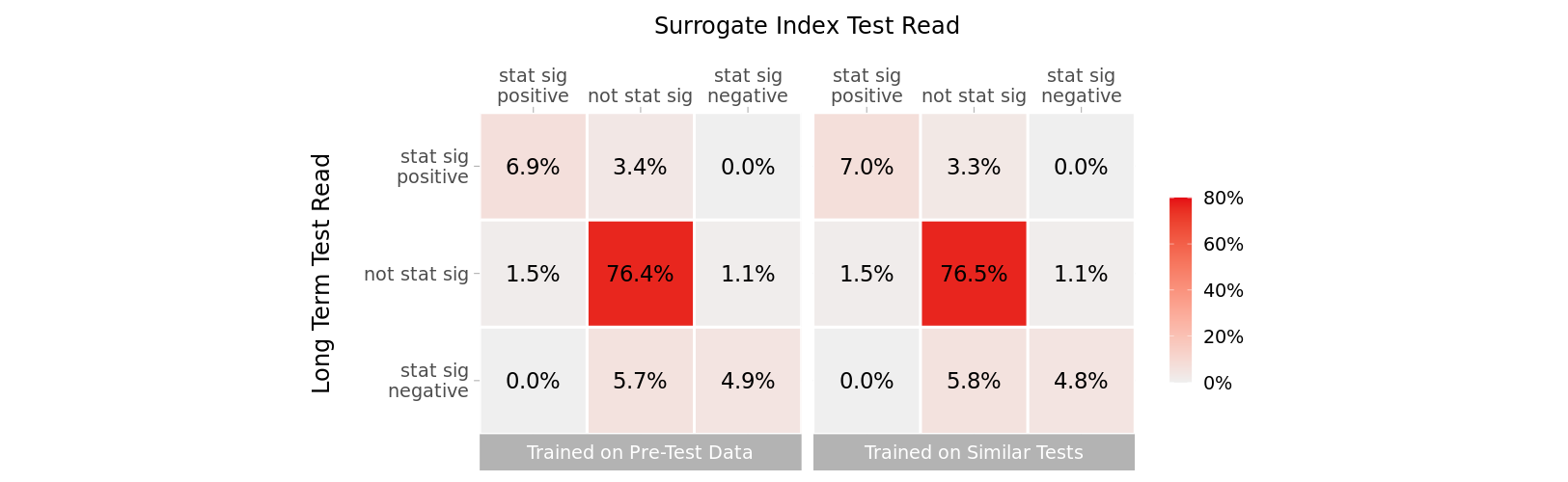}
\caption{Confusion Matrices of the statistical conclusions drawn using the observed 2-month test read (Y-axis) and the surrogate Index test reads (X-axis). The matrix on the left shows the comparison against the surrogate index trained using pre-test data, while the matrix on the right shows the comparison against the surrogate index trained using data from a similar A/B test.}
\end{figure}

At the most extreme, going from a 2-month testing cycle to a 2-week testing cycle could maximally increase total experimentation capacity by 300\%. Admittedly, it’s unlikely that all this extra capacity could be used effectively, but 65\% recall suggests that we would need \textapprox53\% more 2-week experiments to achieve the same gains from 2-month testing cycle, assuming that the ``true" treatment effects distribution doesn't change with the number of experiments run and that the treatment effects of different experiments are additive. Further assuming that there are no or very low marginal costs for additional experiments, this implies that the benefits from increased throughput from a shorter, faster testing cycle likely outweigh the benefits directly measuring long-term effects.

\bibliographystyle{abbrvnat}
\bibliography{lit}

\end{document}